\documentclass[prl,aps,twocolumn]{revtex4}
\usepackage{epsfig}
\usepackage{bm}

\renewcommand{\it}[1]{\textit{#1}}
\newcommand{\Onecol} {\begin{widetext} \onecolumngrid} %% 2 -> 1
\newcommand{\Twocol} {\end{widetext} \twocolumngrid} %% 1 -> 2

\newcommand{\be}{\begin{equation}}
\newcommand{\ba}{\begin{array}}
\newcommand{\bea}{\begin{eqnarray}}
\newcommand{\bfi}{\begin{figure}}
\newcommand{\ee}{\end{equation}}
\newcommand{\ea}{\end{array}}
\newcommand{\eea}{\end{eqnarray}}
\newcommand{\efi}{\end{figure}}

\begin{document}
\title{Shell Model of Two-dimensional Turbulence in Polymer Solutions}
\author{Roberto Benzi$^*$, Nizan Horesh$^\dag$ and Itamar Procaccia$^\dag$}
\affiliation{$^*$ Dipartimento di Fisica and INFM, Universit\`a
``Tor Vergata", Via della Ricerca Scientifica 1, I-00133 Roma, Italy\\
$^\dag$Dept. of Chemical Physics, The
Weizmann Institute of Science, Rehovot,  76100 Israel}
%\pacs{47.27-i, 47.27.Nz, 47.27.Ak}
\begin{abstract}
We address the effect of polymer additives on two dimensional turbulence, an issue
that was studied recently in experiments and direct numerical simulations. We
show that the same simple shell model that reproduced drag reduction in three-dimensional
turbulence reproduces all the reported effects in the two-dimensional case.
The simplicity of the model offers a straightforward understanding of the all
the major effects under consideration.
\vskip 0.2cm
\end{abstract}
\maketitle
%%%%%%%%%%%%%%%%%%%%%%%%%%%%%%%%%%%%%%%%%%%%%%%%%%%%%%%%%%%%%%%%%%%%%%%%%%%%
The effect of soluble polymers on two-dimensional turbulent flows
appear quite surprising \cite{02AK,03BCM}. Instead of reducing the drag as in three-dimensional
turbulence \cite{75Vir,00SW} (thus increasing the large scale velocity 
components \cite{03ACBP,03BDGP,03BCHP}), in two
dimensions polymers appear to
suppress the large scale velocity components. In addition, the polymers affect 
the probability distribution
functions of the velocity field, changing them from sub-gaussian to super-gaussian, with
approximately exponential tails. These phenomena appear generic, and were observed
both in experiments in fast flowing soap films \cite{02AK} and in direct numerical simulation of the
two dimensional viscoelastic equations \cite{03BCM}. In this Letter we argue that these effects
can be understood using a simple shell model of viscoelastic turbulence.

The shell model used to describe the effects of polymers on the turbulent
velocity field had been derived \cite{03BDGP,03BCHP} on the basis of the FENE-P model of 
visco-elastic flows \cite{87BCAH}. It was employed recently to understand successfully drag
reduction in  homogeneous three dimensional visco-elastic turbulence \cite{03BDGP,03BCHP}. It reads
\begin{eqnarray}
\frac{d u_n}{d  t} &=& \frac{i}{3} \Phi_n (u,u) -  \frac{i}{3}
\frac{\nu_p}{\tau} P(B)
\Phi_n (B,B) - \gamma_n u_n + F_n , \nonumber\\
\frac{d B_n}{d  t} &=& \frac{i}{3} \Phi_n (u,B) -  \frac{i}{3}
\Phi_n (B,u) - {1 \over \tau} P(B) B_n - \nu_B k_n^2 B_n,\nonumber\\
P(B) &=& {1 \over 1 - \sum_n B_n^* B_{n} } \ . \nonumber\\
\gamma_n &=&\nu k_n^{4}+\mu k_n^{-4} \ .
\label{SP}
\end{eqnarray}
In these equations $u_n$ and $B_n$ stand for the Fourier amplitudes $u(k_n)$ and $B(k_n)$ of 
the two respective
fields, but as usual in shell models we take $n=0,1,2,\dots$ and the
wavevectors are limited to the set $k_n=2^n$. The nonlinear interaction terms take
the explicit form
\begin{eqnarray}
&&\Phi_n(u,B) = k_n\Big[(1-b) u_{n+2} B^*_{n+1} +(2+b) u^*_{n+1}
B_{n+2}\Big] \nonumber\\&&+ k_{n-1}\Big[(2b+1) u^*_{n-1}B_{n+1}-(1-b)u_{n+1}B^*_{n-1} \Big]\nonumber\\
&&+k_{n-2}\Big[(2+b)u_{n-1}B_{n-2}+(2b+1)u_{n-2}B_{n-1}\Big]\ , \label{Phi}
\end{eqnarray}
with an obvious simplification for $\Phi_n(u,u)$ and
$\Phi_n(B,B)$. Here $b$ is a parameter which can be used to distinguish between
three-dimensional and two-dimensional behavior (and see below for details). 
In accordance with the generalized energy of the FENE-P model,
the non linear terms in our shell model conserves the total energy:
\begin{equation}
E \equiv {1 \over 2} \sum_n |u_n|^2 - {1 \over 2} {\nu_p \over \tau}
\ln \left(1-\sum_n |B_n|^2\right)\ . \label{gener}
\end{equation}
The dissipative term $\gamma u_n$ contains
both a hyper viscosity at the small scales and a friction term at the large scales, to
mimic wall friction in 2-dimensional turbulence. With $\nu_p=0$ the first of
Eqs. \ref{SP} reduces to the well-studied Sabra model of
Newtonian turbulence \cite{98LPPPV}. As in the FENE-P case we
consider $\nu_p/\nu$ to be the concentration of the polymer $c$. For $\nu_p\ne 0$ 
we refer to the model as the SabraP shell model. The forcing in Eqs. (\ref{SP}) is
chosen to input a fixed amount of energy per unit time, i.e.
\begin{equation}
F_{n} = \delta_{n,f} \tilde F/u^*_n\ , \quad \tilde F=10^{-3}\sqrt{2}\quad\text{ for } f=10,11 \ .
\label{forcing}
\end{equation}

\begin{figure}
\centering
\includegraphics[width=.5\textwidth]{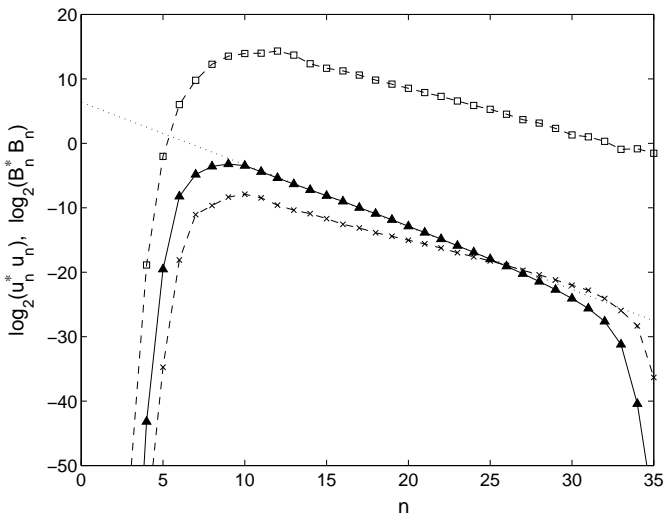}
\caption{Power spectra $S_2(k_n)$ of the SabraP model ($times$) and the Sabra
model (triangles) for $\nu_p=10^{-6}$ and
$\tau=10^6$. Also shown is the spectrum of the $B_n$ field, i.e.
$\langle |B_n|^2\rangle$, which is peaked at the Lumley scale $k_c$.} \label{logspectrum}
\end{figure}

The Sabra model with the parameter $b$ in the range $-1\le b \le 0$ agrees
with the 3-dimensional Navier-Stokes equations in the sense that the energy cascade is normal, from the
large scales to the  smaller scales. For $b<b_c=-1-2^{-2/3}$ the situation 
changes qualitatively, since the energy flux changes direction, going
from smaller to larger scales, as in 2-dimensional turbulence \cite{02GLPP}. 
The role of the direct energy flux is
taken by an enstrophy-like conserved variable $H$ which cascades from larger to
smaller scales:
\begin{equation}
H\equiv \sum_n \left(-\frac{1}{1+b}\right)^n |u_n|^2 \ . \label{enstrophy}
\end{equation}
Thus, if the forcing is applied to an intermediate level $k_f$,
there are two scaling regimes with an inverse energy flux supported on scales
$k_n<k_f$ and a direct enstrophy-like flux for $k>k_f$. The spectrum $S_2(k_n)\equiv \langle u_n
u^*_n\rangle$ has a Kolmogorov exponent$S_2(k_n)\sim k_n^{-2/3}$ in the inverse
cascade regime, whereas in the direct cascade regime the spectrum reads (up to intermittency
corrections) \cite{02GLPP}
\begin{equation}
S_2(k_n)\sim k_n^{-2[1+\log_2(-1/1+b)]/3} \ . \label{inverse}
\end{equation}
When the parameter $b$ is in the range $-1\le b \le 0$ the coupling
to the $B$-field results in drag reduction, as had been discussed in refs. \cite{03BDGP,03BCHP}. 
We demonstrate here that for $b<-1-2^{-2/3}$ the coupling to the $B$-field
suppresses the large scale velocity components. In Fig. \ref{logspectrum}
we show the respective spectra of the Sabra and the SabraP models,
in which the forcing is identical. Note that we force at shells 10 and
11, and therefore we observe only the direct cascade, in agreement with
the experimental and simulational works \cite{02AK,03BCM}. The inverse cascade regime
is destroyed by the large scale friction and is not observable here, as
in \cite{02AK,03BCM}. Eq. (\ref{inverse}) predicts a spectral slope of -0.88 for the present
value of $b$, while the observed slope is -0.96$\pm 0.05$; the difference 
is attributed to the usual intermittency correction. The main point is that
the spectrum with the coupling to the $B$ field exhibits a strong suppression
of the large scales, and the slope reduces to about -0.66$\pm0.05$.
%%%%%%%%%%%%%%%%%%%%%%%%%%%%%%%%%%%%%%
%%%%%%%%%%%%%%%%%%%%%%%%%%%%%%%%%%%%%%%%%%%%%%%%%
We also show, in the same plot, the spectrum of the $B_n$ field,
$\langle |B_n|^2\rangle$. This spectrum has a typical scale $k_c$ which
is known as the Lumley scale \cite{69Lu},
\begin{equation}
\sqrt {S_2(k_c)} k_c \approx \tau^{-1} \ . \label{lumley}
\end{equation}
For wavevector smaller than $k_c$ the $B_n$-spectrum is strongly suppressed, whereas
for $k_n>k_c$ the spectrum is a power law.

The advantage of the present simple model is that the drag enhancement
observed in the spectra lends itself to
a straightforward interpretation. The standard newtonian energy balance equation 
is now changed due to the second term in the generalized energy (\ref{gener}) which
contributes to the
dissipation a positive definite term of the form
$(\nu_p/\tau^2)P^2(B) \sum_n|B_n|^2$. Thus the energy balance
equation reads
\begin{equation}
\tilde F = \epsilon+(\nu_p/\tau^2)P^2(B) \sum_n|B_n|^2 +\mu\sum_nk_n^{-4} |u_n|^2 \ , \label{enbal}
\end{equation}
where 
\begin{equation}
\epsilon = \nu \sum_nk^4_n |u_n|^2
\end{equation} 
is the dissipation due to hyper viscosity. In 3-dimensional turbulence the dissipation is
independent of the viscosity $\nu$ (This fact is known in the jargon as ``the viscous anomaly"). In the 
present case where the energy flux reverses direction $\epsilon$ vanishes
in the limit $\nu\to 0$, and for large Reynolds numbers can be neglected \cite{02BCMV}.
Observing the fact that the last sum in (\ref{enbal}) is strongly
dominated by the forcing scale $k_f$ we can safely estimate
\begin{equation}
|u_f|^2\approx \frac{Ck_f^4}{\mu} \left[\tilde F -(\nu_p/\tau^2)P^2(B) \sum_n|B_n|^2\right] \ ,
\label{uf}
\end{equation}
where $C$ is a constant of the order of unity. The first term in the parenthesis represents
the Sabra value of $|u_f|^2$. The subtraction of the positive definite
second term is the explanation of the observed spectra in Fig. \ref{logspectrum}.
%%%%%%%%%%%%%%%%%
\begin{figure}
\centering
\includegraphics[width=.5\textwidth]{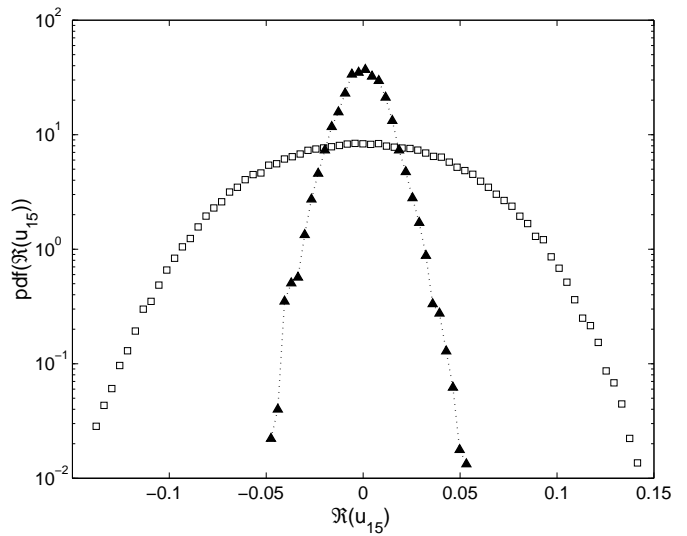}
\caption{Probability distribution function of $\Re(u_{15})$ for the Sabra (squares)
and the SabraP (triangles) models. The forcing is shown in Eq. (\ref{forcing}).} \label{velfluc}
\end{figure}
%%%%%%%%%%%%%%%%%%%%%%

\begin{figure}
\centering
\includegraphics[width=.5\textwidth]{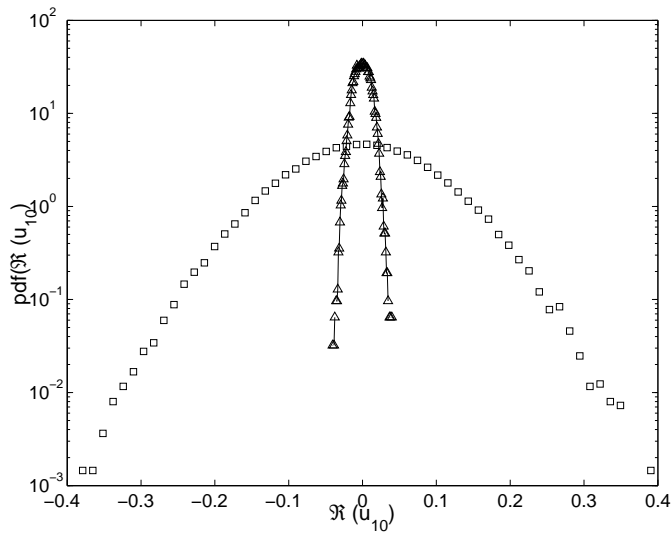}
\caption{Probability distribution function of $\Re(u_{10})$ with Gaussian forcing
on level 10, 11. Symbols are as in Fig. \ref{velfluc}. Fitting a stretched 
exponential tails (cf. Eq. (\ref{tail}) we find exponents $2 \pm 0.15$ and $1.59 \pm 0.2$
for the Sabra and SabraP models respectively. } \label{velflucg}
\end{figure}

\begin{figure}
\centering
\includegraphics[width=.5\textwidth]{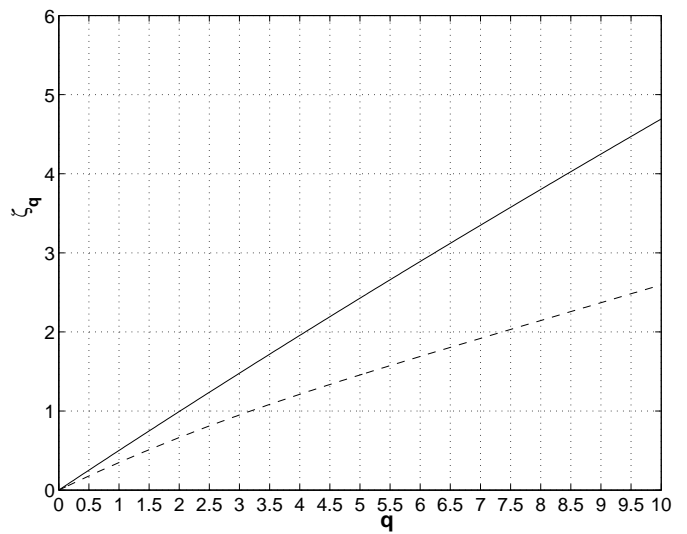}
\caption{Scaling exponents $\zeta_q$ for the Sabra (solid line) and
SabraP (dashed line) models. The increased intermittency in the SabraP
model is exhibited by the increased nonlinearity of $\zeta_q$ as a function of $q$.}
 \label{scalexp}
\end{figure}

Another phenomenon that was discovered in the experiments and the direct
numerical simulations is a significant change in the pdf's of the velocity
fluctuations. We demonstrate that the same phenomenon is recaptured
in our SabraP model, showing that it is generic to the interaction
with the $B$ field. In Fig. \ref{velfluc} we present the probability
distribution functions (pdf) of $u_n$ for $n=15$, which is within the 
bulk of the direct enstrophy cascade. 
The qualitative change in the pdf is obvious. We quantify the difference by fitting the tails of the
pdf's to stretched exponential forms
\begin{equation}
p[\Re(u_n)]\sim \exp(-\beta [\Re(u_n)]^{-\alpha_n}) \quad\text{for}~ \Re(u_n)> \sigma_n \ , \label{tail}
\end{equation}
where $\sigma_n\equiv \sqrt{\langle|u_n|^2\rangle}$. We find $\alpha_{15}\approx 3.45\pm 0.15$
and $\alpha_{15}\approx 1.85\pm 0.2$ for
the Sabra and SabraP models respectively.  We note that in \cite{03BCM} the forcing
was Gaussian. We repeated our simulations for a Gaussian force and found results
in correspondence with \cite{03BCM}, see Fig. \ref{velflucg}. 
%%%%%%%%%%%%%%%%%
%%%%%%%%%%%%%%%%%%%%%%
The change in pdf's indicates that the coupling 
to the $B$-field results in increased intermittency. This can be quantified by measuring
the scaling exponents of the standard structure functions $S_q(k_n)$,
\begin{equation}
S_q(k_n) \equiv \langle |u_n|^q \rangle \sim k_n^{\zeta_q} \ .
\end{equation}
Indeed, the scaling exponents of the SabraP models are significantly more
nonlinear (as a function of $q$) then the corresponding exponents of the Sabra model, 
cf,  Fig. \ref{scalexp}.
%%%%%%%%%%%%%%%%%
%%%%%%%%%%%%%%%%%%%%%%

\begin{figure}
\centering
\includegraphics[width=.5\textwidth]{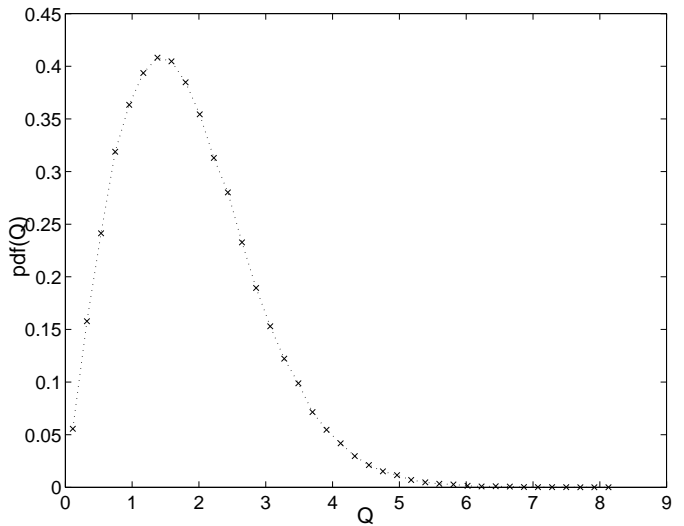}
\caption{Probability distribution function of the quantity $Q$, see text for details.}
 \label{Qpdf}
\end{figure}

The simplicity of our model again allows us to offer a qualitative understanding of the increased
intermittency. Focus on the Lumley scale $k_c$, Eq. \ref{lumley}.
We denote  $u_c=u(k_c)$ and $B_c = B(k_c)$. The total ``energy" of the B
field will be denoted by $Q$ where $Q = \sum_n |B_n|^2$. In light of the spectrum 
shown in Fig. \ref{logspectrum} we can estimate $Q\approx |B_c|^2$ up to coefficients
of the order of unity. The flux of energy from the flow to the polymer is proportional to 
$k_c Q u_c$. cf. Eq. \ref{Phi}. Therefore we can write an approximate equation, in which
we disregard the difference between complex numbers and their moduli:
\begin{equation}
\frac{dQ}{dt} \approx  -\frac{Q}{\tau} + Q k_c u_c \ . \label{argu}
\end{equation}
From this equation we see that, because $k_c u_c \approx 1/\tau$, that the fluctuations
of Q are expected to be very strong. Moreover, even if $u_c$ in Eq. (\ref{argu})
were a Gaussian process, $Q$ would turn out to be log-normally distributed, due
to the multiplicative role of $u_c$ in Eq. (\ref{argu}). Indeed, a direct
measurement of the distribution of $Q$, see Fig. \ref{Qpdf}, confirms the existence
of the anomalously long tail in this pdf. Finally, due to Eq. (\ref{uf}), we see
that an anomalous tail in $Q$ would directly induce an anomalous tail in the pdf
of $u_f$, as seen in the simulations.
%%%%%%%%%%%%%%%%%
%%%%%%%%%%%%%%%%%%%%%%
Finally, it is interesting to ask whether in 2-dimensional flows we can have {\em enstrophy}
drag reduction, since this quadratic invariant replaces the energy as the quantity cascaded
to smaller scales. In our model this is impossible; our forcing with constant energy flux
also produces a constant enstrophy flux, and as can be seen from Eq. (\ref{enstrophy}),
the enstrophy is dominated by the larger scales (small $k_n$). Thus the reduction in energy
is accompanied with the reduction of enstrophy. To switch the role of the scales to get
the enstrophy to be dominated by the small scales (large $k_n$) requires changing $b$ to
compensate for the decrease of $|u_n|^2$ with increasing $k_n$. It is easy to check however
that the change of roles occurs precisely at the value of $b$ where the inverse cascade
of energy disappears, i.e. $b=-1-2^{-2/3}$. We thus cannot have enstrophy drag reduction
in this model, and in our opinion neither in the 2-dimensional Navier-Stokes case.

In summary, we conclude that the results observed for the effects of dilute polymers
on 2-dimensional turbulence, as observed in experiments and direct numerical
simulations, can be readily understood with a simple
model of viscoelastic flows once the cascade of energy is inverted. 
The same model exhibits drag reduction when the parameters allow a direct energy
cascade. These observations lend further support for the use of shell models
as simple and transparent tools for understanding a variety of turbulent phenomena
that remain more obscure when the full hydrodynamics equations are employed.

%%%%%%%%%%%%%%%%%%%%%%%%%%%%%%%%%
\acknowledgments \vskip 0.5 cm This work was supported in part by
the European Commission under a TMR grant, the Minerva
Foundation, Munich, Germany, and the US-Israel Binational Science Foundation.
%%%%%%%%%%%%%%%%%%%%%%%%%%%%%%%%%%


\begin{thebibliography}{99}

\bibitem{02AK}
Y. Amarouchene and H. Kellay, Phys. Rev. lett. {\bf 89} 104502-1 (2002)

\bibitem{03BCM}
G. Boffetta, A. Celani and S. Musacchio, arXiv:nlin.CD/0303008v1

\bibitem{75Vir}
P.S. Virk, AIChE J. {\bf 21}, 625 (1975); Nature {\bf 253}, 109 (1975).

\bibitem{00SW}

K. R. Sreenivasan and C. M. White, J. Fluid Mech. {\bf 409}, 149 (2000).

\bibitem{03ACBP}
E. de Angelis, C. Casciola, R. Benzi, and R. Piva, Phys. of Fluids, submitted.

\bibitem{03BDGP}
R. Benzi, E. De Angelis, R. Govindarajan and I. Procaccia, 
``Shell Model for Drag Reduction with Polymer
Additive in Homogeneous Turbulence", Phys. Rev. E, submitted.

\bibitem{03BCHP}
R. Benzi, E. S.C. Ching, N. Horesh and I.r Procaccia, ``Theory of concentration dependence in drag
reduction by polymers and of the MDR asymptote", Phys. Rev. Lett, submitted.

\bibitem{87BCAH}
R.B. Bird, C.F. Curtiss, R.C. Armstrong and O. Hassager, {\it Dynamics of Polymeric Fluids}
Vol.2 (Wiley, NY 1987).

\bibitem{98LPPPV}
V.S. L'vov, E. Podivilov, A. Pomyalov, I. Procaccia and D.
Vandembroucq., Phys. Rev. E {\bf 58} 1811 (1998).

\bibitem{02GLPP}
T. Gilbert, V. S. L'vov, A. Pomyalov, and I. Procaccia, Phys. Rev. Lett., {\bf 89}, 074501 (2002). 


\bibitem{69Lu}
J. L. Lumley, Ann. Rev. Fluid Mech. {\bf 1}, 367 (1969).

\bibitem{02BCMV}
G. Boffetta, A. Celani, S. Musacchio and M. Vergassola, Phys. Rev. E, {\bf 66}, 026304 (2002).


\end{thebibliography}
\end{document}